\documentclass[twocolumn,eqsecnum,floats,aps,nofootinbib,showpacs]{revtex4}
\usepackage{amsmath,amsthm,amssymb,amsfonts}
\usepackage{fontenc,dsfont,layout}
\usepackage[dvips]{graphicx}

\newcommand{\bra}[1]{|{#1}\rangle}                        
\newcommand{\ket}[1]{\langle{#1}}

\newcommand{\bazad}[2]{^{\textrm{\tiny{0}}} \! \omega^{#1}_{#2}}

\begin{document}

\title{Loop Quantum Cosmology of Diagonal Bianchi Type I model: simplifications and scaling problems}

\author{{\L}ukasz Szulc}
\email{lszulc@fuw.edu.pl}
\affiliation{Institute of Theoretical Physics, University of Warsaw ul. Ho\.{z}a 69, 00-681 Warszawa, Poland}

\begin{abstract}
A simplified theory of the diagonal Bianchi type I model coupled with a massless scalar field in loop quantum cosmology is constructed according to the $\bar{\mu}$ scheme. Kinematical and physical sectors of the theory are under good analytical control as well as the scalar constraint operator. Although it is possible to compute numerically the nonsingular evolution of the three gravitational degrees of freedom, the naive implementation of the $\bar{\mu}$ scheme to the diagonal Bianchi type I model is problematic. The lack of the full invariance of the theory with respect to the fiducial cell and fiducial metric scaling causes serious problems in the semiclassical limit of the theory. Because of this behavior it is very difficult to extract reasonable physics from the model. The weaknesses of the implementation of the $\bar{\mu}$ scheme to the Bianchi I model do not imply limitations of the $\bar{\mu}$ scheme in the isotropic case.
\end{abstract}

\maketitle

\section{Introduction}
During the last years loop quantum cosmology \cite{Bojo-rev} (LQC) has become an attractive area of research for cosmology and quantum gravity community. The loop quantum gravity \cite{Thiemann, Rovelli, AL-rev} (LQG) which inspired quantization\footnote{To learn more about deep relation between LQG and LQC see introductory review \cite{A-intro}.} of the symmetry reduced cosmological models, allows one to address fundamental questions about the fate of classical singularity and quantum gravitational corrections in the early universe. While the topic of the isotropic and homogeneous sector of the LQC originated in \cite{Bojo-iso} is well understood \cite{ABL,APS,Van-open,APVS,SKL,Szulc-open,KLS-obs} there still is work to be done in the homogeneous, but nonisotropic sector. Although loop quantum dynamics is not fully understood in this sector, already the first calculations in the quantum homogeneous models \cite{Bojo-BI,BIX} suggest a completely different structure of the space-time near classical singularities. After the discovery of the so-called "improved scheme" in \cite{APS}, analytical issues of kinematics and dynamics were studied in detail in the case of the simplest homogeneous diagonal Bianchi type I (BI) model \cite{5}. Although there has been recent progress in the LQC diagonal BI model at the level of effective equations of motion \cite{1,2,3}, the predictions coming directly from the quantum theory are still missing. Potential numerical simulations of the quantum dynamics in the theory constructed in \cite{5} are more difficult than in the isotropic case \cite{APS} because of the complexity of the quantum scalar constraint. While the numerical study of the \cite{5} model is still to be done, we can in the meantime focus our attention on recent papers \cite{KL-self,4}. In \cite{4} it was shown that a small simplification in the quantum theory leads to an exactly soluble LQC model. Similar simplifications enable us to prove self-adjointness of the quantum Hamiltonian in a purely analytical manner \cite{KL-self}. This motivates us to investigate similar simplifications in the LQC diagonal Bianchi I model. 

Similar results have also been obtained independently by the Madrid group \cite{MMP,MGM}.

Furthermore, so far in the literature there are two kinds of "loop regularization" of the gravitational part of the scalar constraint for the BI models described in \cite{3}, the so-called $\bar{\mu}$ and $\bar{\mu}^{\prime}$. The $\bar{\mu}^{\prime}$ scheme has better scaling properties (see \cite{3}); however, a quantum theory for this scheme is much more difficult to construct. In this paper we study the $\bar{\mu}$ scheme quantization described in \cite{5} and simplify it to some extent. Unfortunately, such a theory has a limited domain of applicability. Full scaling invariance with respect to the different choices of the fiducial cell is broken, which causes fiducial cell and fiducial metric dependence on the semiclassical limit for the fiducial cells (and fiducial metric) different from cubical (see Sec. \ref{3b} for a more detailed discussion). These problems are especially important for the case of noncompact topology of the three-dimensional spatial slice $\Sigma$ of the foliation, where different shapes of the fiducial cells (and fiducial metrics different than isotropic) are allowed. The quantum theory should be invariant with respect to changes to different fiducial cells (and fiducial metrics) just like the classical theory does respect this invariance. The model here does not respect the fiducial invariance, resulting in the semiclassical limit of the theory is being unfortunately ill defined.

This paper is organized as follows. In Sec. II the classical theory of BI cosmology is described in terms of Ashtekar variables. In Sec. III we consider the quantum theory of BI in the volume and the connection "($\eta$)" representations. Section IV describes a unitary transformation $W$ which allows one to use continuous 3D Fourier transform in order to extract semiclassical states from the theory. Section V contains a discussion. In the appendix the reader can find numerical strategy used in the simulations.

\section{Classical Theory}
The scalar constraint for the (minimally coupled) general relativity with a massless scalar field in terms of Ashtekar variables is given by
\begin{align}\label{total-sc}
C'&=\frac{1}{16 \pi G} \int_{\Sigma} d^3 x N(x)\Big( e^{-1}E^a_i
  E^b_j {\varepsilon^{ij}}_k F_{ab}^k \nonumber \\&-
  2(1+\gamma^2) e^{-1}E^a_i E^b_j K^i_{[a}
  K^j_{b]} \Big) + C_{\phi},
\end{align}
where ${\rm det} E = e^2$ and $K^i_a$ stands for extrinsic curvature of the spatial slice $\Sigma$. Symplectic structure is defined by the following Poisson bracket: $$\{A(x)^i_a, E(y)^b_j \} = 8 \pi G \gamma \delta^i_j \delta^b_a \delta^3(x,y) \ .$$ Spatial part of the metric tensor for the diagonal Bianchi I models is given by
\begin{equation}\label{metric}
q_{ab}= a^2_1(t) \ \bazad{1}{a} \ \bazad{1}{b} + a^2_2(t) \ \bazad{2}{a} \ \bazad{2}{b} + a^2_3(t) \ \bazad{3}{a} \ \bazad{3}{b} \ ,
\end{equation}
where the left invariant one-forms satisfy $\partial_{[a} \ \bazad{i}{b]}=0$ (i.e. structure constants $C^i_{jk}=0$). If we want to pass to the Hamiltonian formulation for our BI model we must be careful, because the scalar constraint (\ref{total-sc}) for the metric (\ref{metric}) is infinite. In order to make the integration over spatial slice finite we can replace $\Sigma \to \mathcal{V}$ as 
\begin{equation}
\int_{\Sigma} \to \int_{\mathcal{V}} \ ,
\end{equation}
where $\mathcal{V}$ is a finite cell or, in the case of Bianchi I, we can choose topology of $\Sigma$ to be a three-dimensional torus. 
The metric tensor given by the (\ref{metric}) reduced symplectic structure in terms of Ashtekar variables is defined as $\{ \tilde{c}^i, \tilde{p}_j \}=\delta_j^i 8 \pi G \gamma V_0^{-1}$, where $\tilde{c}^i \sim \gamma \dot{a}_i$ and $\tilde{p}_i = \pm a_j a_k$ \footnote{Plus and minus mean two possible orientations of the spatial triads $e^i_a = a(t)_{(i)} \ \bazad{i}{a}$} for $\varepsilon_{ijk}=1$. $V_0 = L_1 L_2 L_3$ denotes the fiducial volume of the finite cell (or a volume of the 3-torus). After symmetry reduction given by the metric (\ref{metric}) constraint (\ref{total-sc}) is reduced to the following one
\begin{align}\label{constr}
C'&=-N(t') \frac{1}{8 \pi G\gamma^2} \frac{1}{\sqrt{|p_1 p_2 p_3|}} (c^1 p_1 c^2 p_2 + c^1 p_1 c^3 p_3 \nonumber \\&+ c^2 p_2 c^3 p_3) + N(t')\frac{1}{2} \frac{\Pi^2_{\phi}}{\sqrt{|p_1 p_2 p_3|}}=0 \ .
\end{align}
The variables $c^i$ and $p_j$ are now rescaled as follows
\begin{equation}\label{resc}
c^{1} = L_1 \tilde{c}^1 \quad p_1 = \tilde{p}_1 L_2 L_3 = \pm L_2 a_2 \ L_3 a_3 \ .
\end{equation}
The symplectic structure is now given by $$ \{ c^i, p_j \}=\delta_j^i 8 \pi G \gamma .$$ Moreover for the scalar field in (\ref{constr}) we have $\{\phi, \Pi_{\phi} \}=1$. If one chooses the lapse function to be $N(t')|p_1 p_2 p_3|^{-1/2}=1$ the scalar constraint becomes
\begin{equation}\label{reg1}
C=-\frac{1}{8 \pi G\gamma^2} (c^1 p_1 c^2 p_2 + c^1 p_1 c^3 p_3 + c^2 p_2 c^3 p_3)+ \frac{1}{2}\Pi^2_{\phi} = 0 \ .
\end{equation}
Hamiltonian equations of motion \footnote{A time derivative in (\ref{motion}) is taken with respect to the time $t$ associated with laps $N(t)=|p_1 p_2 p_3|^{1/2}$}
\begin{align}\label{motion}
&\dot{c}^i = \{ c^i, C\} \quad \dot{p}_j = \{ p_j, C \} \\
&\dot{\Pi}_{\phi} = \{\Pi_{\phi}, C \} \quad \dot{\phi} = \{ \phi, C \} \nonumber 
\end{align}
give us the following conditions:
\begin{equation}
c^{(i)} p_i ={\rm const}_i \quad \Pi_{\phi} = {\rm const}_{\phi} \ ,
\end{equation}
where $(i)$ means no sum over $i$. Let us set ${\rm const}_i := 8 \pi G \gamma \hbar \mathcal{K}_i$ and ${\rm const}_{\phi} := \hbar \sqrt{8\pi G} \mathcal{K}_{\phi}$. The solutions can be written as functions $p_i = p(\phi)_i$ given by
\begin{equation}\label{p-phi}
p_i (\phi) = p_{0i} \exp\big(\sqrt{8\pi G} \frac{1- \kappa_i}{\kappa_{\phi}} (\phi - \phi_0) \big) \ ,
\end{equation}
where $\mathcal{K}_i= \mathcal{K} \kappa_i$ and $\mathcal{K}_{\phi}=\mathcal{K} \kappa_{\phi}$ are rescaled such that
\begin{align}\label{kappa}
& \kappa_1 + \kappa_2 + \kappa_3 = 1 \\
& \kappa_1^2 + \kappa_2^2 + \kappa_3^2 + \kappa^2_{\phi} =1 \nonumber .
\end{align}
There are two types of solutions: "Kasner-like", when two $\kappa_i$ are positive and the other is negative, and "Kasner-unlike," when all three $\kappa_i$ are positive. Let us now define, for the purpose of the quantum theory, a new symplectic structure as
\begin{equation}
\{ \eta^i, V_j \}=12 \pi G \gamma \delta_j^i \ ,
\end{equation}
where new variables are defined as $$\eta^i =\frac{c^i}{\sqrt{|p_{(i)}|}}, \quad  V_i={\rm sgn} ({p_i})|p_i|^{3/2}  .$$
The classical solutions (\ref{p-phi}) in terms of new variables $V_i$ are in the form
\begin{equation}\label{V-phi}
V_i (\phi) = V_{0i} \exp\big(\pm \sqrt{8\pi G} \frac{3}{2} \frac{(1- \kappa_i)}{|\kappa_{\phi}|} (\phi - \phi_0) \big) \ ,
\end{equation}
where we put $\kappa_\phi = \pm |\kappa_{\phi}|$, because $\kappa_{\phi}$ can be positive or negative. The total physical volume of the fiducial cell (or a 3-Torus) is defined as $V=|V_1 V_2 V_3|^{1/3}$. 

\section{Quantum Theory}
\subsection{Kinematics - Volume representation}
A quantum theory in the improved (or not improved, so-called ${}^{o}\mu$) scheme was constructed in detail in \cite{5}. In this subsection we briefly recall the kinematics of the diagonal Bianchi I model in the so-called $\bar{\mu}$ scheme. The kinematical Hilbert space is given by $\mathcal{H}_{\rm Kin}=L^2(\mathbb{R}_{\rm{Bohr}}, d\mu_{\rm {Bohr}})^{\otimes 3}$ with the orthonormal basis elements labeled by three real numbers $$\bra{\nu_1,\nu_2, \nu_3}:=\bra{\nu_1} \otimes \bra{\nu_2} \otimes \bra{\nu_3} \ ,$$ where $\mathbb{R}_{\rm{Bohr}}$ stands for the Bohr compactification of a real line. The kinematical scalar product is defined as $$\ket{\nu_1,\nu_2, \nu_3} \bra{\tilde{\nu}_1,\tilde{\nu}_2, \tilde{\nu}_3}= \delta_{\nu_1, \tilde{\nu}_1} \delta_{\nu_2, \tilde{\nu}_2} \delta_{\nu_3, \tilde{\nu}_3}.$$
Any state $\bra{\psi} \in \mathcal{H}_{\rm Kin}$ can be decomposed in the orthonormal basis as $$\bra{\psi} = \sum_{\nu_1} \sum_{\nu_2} \sum_{\nu_3} \psi(\nu_1, \nu_2, \nu_3) \bra{\nu_1,\nu_2, \nu_3}$$ with the norm $N$ given by $$N^2=\sum_{\nu_1} \sum_{\nu_2} \sum_{\nu_3} \bar{\psi}(\nu_1, \nu_2, \nu_3) \psi(\nu_1, \nu_2, \nu_3) .$$ There exists a volume operator defined as
\begin{equation}
\hat{V}\bra{\nu_1,\nu_2, \nu_3}=6 \pi \sqrt{\Delta} \gamma l_{\rm Pl}^2 |\nu_1 \nu_2 \nu_3|^{1/3} \bra{\nu_1,\nu_2, \nu_3} ,
\end{equation}
which is constructed from the components of the volume operator as $\hat{V} = |\hat{V}_1 \hat{V}_2 \hat{V}_3|^{1/3}$. The constant $\Delta$ will be determined later. Each of the $\hat{V}_i$ operators is given by
\begin{equation}
\hat{V}_i \bra{\nu_1,\nu_2, \nu_3}=6 \pi \sqrt{\Delta} \gamma l_{\rm Pl}^2 \nu_i \bra{\nu_1,\nu_2, \nu_3}
\end{equation}
for $i=1,2,3$. The operators corresponding to the connection components $c_i$ do not exist in $\mathcal{H}_{\rm Kin}$, but there are important unitary shift operators (for $i=1,2,3$) 
\begin{equation}\label{shift1}
\hat{U}^{(i)}_{(b)} = \widehat{\exp\big(i b\frac{\bar{\mu}_{(i)} c^i}{2}\big)} \ ,
\end{equation}
where $b \in \mathbb{R}$ (and there is no sum over $i$). Their action on a basis element is defined as follows
\begin{equation}\label{shift}
\hat{U}^{(2)}_{(b)} \bra{\nu_1,\nu_2, \nu_3}= \widehat{\exp\big(ib\frac{\bar{\mu}_2 c^2}{2}\big) }\bra{\nu_1,\nu_2, \nu_3} = \bra{\nu_1,\nu_2 +b, \nu_3} \nonumber .
\end{equation}
\subsection{Dynamics - Volume representation} \label{3b}
In the paper \cite{5} one can find a detailed construction of the operator corresponding to the (\ref{constr}) in the full Loop Quantum Gravity scheme. The full LQG quantization of the scalar constraint (\ref{total-sc}) gives us quantum corrections to the term $ e^{-1}E^a_i E^b_j {\varepsilon^{ij}}_k $ by the so-called Thiemann trick. It also gives quantum corrections to the term $F_{ab}^k$ by SU$(2)$ holonomy along a suitable loop (see Eqs. (5.2) and (5.4) in \cite{5} for the case of the diagonal Bianchi I model). However, the simplified LQC model proposed in \cite{4} can be interpreted as a quantum model, where only the LQG effects coming from curvature 2-from $F_{ab}^k$ are taken into account.
In a similar way we construct in this section an operator corresponding to (\ref{reg1}) where only a curvature 2-form $F_{ab}^k \sim c^{(i)} c^{(j)} {\varepsilon_{ij}^k}$ is loop quantized, so in order to simplify quantum theory the Thiemann trick is simply ignored. Therefore, we can use a regularized version of (\ref{reg1}) (described in \cite{1,3}) given by
\begin{align}\label{reg}
C_{\rm reg}=&-\frac{1}{8 \pi G\gamma^2} \left( \frac{\sin{\bar{\mu}_1 c^1}}{\bar{\mu}_1} p_1 \frac{\sin{\bar{\mu}_2c^2}}{\bar{\mu}_2} p_2  \right. \nonumber \\ &+ \left. \frac{\sin{\bar{\mu}_1 c^1}}{\bar{\mu}_1} p_1 \frac{\sin{\bar{\mu}_3 c^3}}{\bar{\mu}_3} p_3 + \frac{\sin{\bar{\mu}_2 c^2}}{\bar{\mu}_2} p_2 \frac{\sin{\bar{\mu}_3 c^3}}{\bar{\mu}_3} p_3\right) \nonumber \\ &+ \frac{1}{2} \Pi_{\phi}^2=0 \ ,
\end{align}
where
\begin{equation} \label{mu bar}
\bar{\mu}_1= \frac{\sqrt{\Delta}}{\sqrt{|p_1|}}, \quad \bar{\mu}_2= \frac{\sqrt{\Delta}}{\sqrt{|p_2|}}, \quad \bar{\mu}_3= \frac{\sqrt{\Delta}}{\sqrt{|p_3|}} \ . 
\end{equation}
$\Delta$ stands for the minimal non-zero eigenvalue of the area operator \cite{AL-area}. Let us now investigate the last two expressions in more detail. The (\ref{reg}) was obtained in two steps. Firstly the fiducial cell $V_0=L_1 L_2 L_3$ was introduced in order to make the scalar constraint (\ref{total-sc}) finite. Secondly the loop regularization with the condition (\ref{mu bar}) was used to obtain the formula (\ref{reg}). Because there is no preferred fiducial cell $\mathcal{V}$ the theory should not depend on the choice of the fiducial cell. In other words the two different cells should give the same physical evolution. This indeed is the case for the isotropic $k=0$ model \cite{APS}. However, for the quantum model determined by (\ref{reg}) the situation is different. It can be easily seen from the (\ref{reg}), namely the expression $\sin({\bar{\mu}_1 c^1})$ (the scaling properties for the ${\bar{\mu}_2 c^2}$ and ${\bar{\mu}_3 c^3}$ are analogous). Let us notice that 
\begin{equation} \label{mu1}
\bar{\mu}_1 c^1 = \frac{\sqrt{\Delta} c^1}{|p_1|^{1/2}} \sim \frac{\dot{a}_1}{\sqrt{a_2 a_3}} \ .
\end{equation}
Now, if we rescale our fiducial cell as $$V_0 = L_1 L_2 L_3 \to V_0^{\prime}= L_1 l_1 L_2 l_2 L_3 l_3 = l_1 l_2 l_3 V_0 \ ,$$ the (\ref{mu1}) is rescaled according to the (\ref{resc}) as
\begin{equation}\label{limitat}
\bar{\mu}_1 c^1 \to \bar{\mu}_1 c^1 \frac{l_1}{\sqrt{l_2 l_3}} \ .
\end{equation}
The expression $\bar{\mu}_1 c^1$ is then not invariant under the change of the fiducial cell. The $\sin({\bar{\mu}_1 c^1})$ is promoted to the shift operator by (\ref{shift1}), so it seems that the magnitude of the volume shift in the operator corresponding to the (\ref{reg}) is fiducial scale dependent. This property can be summarized as follows: The classical dynamics does not depend on the choice of the fiducial cell, while the quantum dynamics does. Such behavior is very unfortunate, because it can generate several problems as in the old isotropic LQC models. One can argue that for the 3-Torus topology where the total coordinate volume of the spatial slice $\Sigma$ is fixed, the $\bar{\mu}$ condition has correct implementation \cite{MGM,MMP}. However, in the case of noncompact $\Sigma$ topology there is no doubt that (\ref{limitat}) is a serious defect. So what can we do about it? There are at least two possibilities. We can develop a better loop regularization with correct scaling properties or we can assume that the only consistent (partial) solution of the rescaling problem is to use only cubical fiducial cells. Then for the cubical choice we have $l_i = l$ ($i=1,2,3$) and only for the rescaling defined by $V_0 \to l^3 V_0$ (each $a_i \to la_i $) the (\ref{limitat}) and the volume shift defined by $\bar{\mu}_i c^i$ ($i=1,2,3$) are invariant. In the rest of this paper the cubical shape assumption of the fiducial cell is made. Moreover, also the fiducial metric $^{\textrm{\tiny{0}}} q_{ab} = \ \bazad{1}{a} \ \bazad{1}{b} + \bazad{2}{a} \ \bazad{2}{b} + \bazad{3}{a} \  \bazad{3}{b}$ must be assumed to be isotropic.
  
Our task now is to promote (\ref{reg}) to a well-defined, symmetric operator. Let us start with a matter term $(1/2)\Pi_{\phi}^2$. Because we are interested only in quantum gravitational effects the scalar field $\phi$ is quantized as usual $$\hat{\phi} \psi = \phi \psi \quad \hat{\Pi}_{\phi}\psi=-i\hbar \partial_{\phi} \psi ,$$ where $\psi$ belongs to the total kinematical Hilbert space defined as $\mathcal{H}_{\rm tot}=\mathcal{H}_{\rm Kin} \otimes L^2(\mathbb{R}, d\phi)$. The next step is to use kinematical tools described in the previous section to get an operator corresponding to the function $\sin\bar{\mu}_{(i)} c^i$ defined as follows
\begin{align}
&\widehat{\sin(\bar{\mu}_{(k)} c^k) }\bra{\nu_k} = \frac{1}{2i}\Big(\widehat{ \exp(i\bar{\mu}_{(k)} c^k )}- \nonumber \\  & \widehat{\exp(-i\bar{\mu}_{(k)} c^k ) }\Big) \bra{\nu_k} = \frac{1}{2i}(\bra{\nu_k + 2} - \bra{\nu_k - 2})
\end{align}
for $k=1,2,3$. Let us now fix the factor ordering ambiguity of the (\ref{reg}) operator in a following way. Define now an operator
\begin{align}\label{actionO}
\hat{\underline{O}}_k \bra{\nu_k} :&=\frac{1}{2}\Big(\frac{\widehat{\sin\bar{\mu}_{(k)} c^k}}{\sqrt{\Delta}} \hat{V}_k + \hat{V}_k \frac{\widehat{\sin\bar{\mu}_{(k)} c^k}}{\sqrt{\Delta}}\Big) \bra{\nu_k}  \\ &= \frac{6\pi \gamma l_{\rm Pl}^2}{2i} \big( (\nu_k + 1) \bra{\nu_k+2} - (\nu_k -1) \bra{\nu_k -2} \big) \nonumber
\end{align}
and notice the following property: $[\hat{\underline{O}}_i, \hat{\underline{O}}_j] \bra{\nu_1, \nu_2,\nu_3}=0$ for $i\neq j$. In terms of (\ref{actionO}) operators we get a symmetric (with respect to the kinematical scalar product) gravitational part of the scalar constraint operator given by 
\begin{align}\label{scalar vol}
\hat{C}_{\rm gr} \bra{\nu_1,\nu_2, \nu_3} &=-\frac{1}{8 \pi G \gamma^2}\Big( \hat{\underline{O}}_1 \hat{\underline{O}}_2 +\hat{\underline{O}}_2 \hat{\underline{O}}_3\nonumber \\ & +\hat{\underline{O}}_1 \hat{\underline{O}}_3 \Big) \bra{\nu_1,\nu_2, \nu_3} \ .
\end{align}
The action of the (\ref{scalar vol}) on a basis element is as follows\footnote{An operator (\ref{diffC}) can be obtained also for the laps function $N(t)=1$ by neglecting the inverse triad corrections and by specific choice of the factor ordering in the full quantum constraint.} 
\begin{align}\label{diffC}
\hat{C}_{\rm gr} \bra{\nu_1,\nu_2, \nu_3} &= \frac{9}{8} \pi G \hbar^2 \big[ (\nu_1+1)(\nu_2+1)\bra{\nu_1+2, \nu_2+2, \nu_3}  \nonumber \\
&-(\nu_1-1)(\nu_2+1)\bra{\nu_1-2, \nu_2+2, \nu_3} \nonumber \\ 
&-(\nu_1+1)(\nu_2-1) \bra{\nu_1+2, \nu_2-2, \nu_3} \nonumber \\
&+(\nu_1-1)(\nu_2-1)\bra{\nu_1-2, \nu_2-2, \nu_3} \nonumber \\
&+(\nu_2+1)(\nu_3+1)\bra{\nu_1, \nu_2+2, \nu_3+2} \nonumber \\ 
&-(\nu_2-1)(\nu_3+1)\bra{\nu_1, \nu_2-2, \nu_3+2} \nonumber \\ 
&-(\nu_2+1)(\nu_3-1)\bra{\nu_1, \nu_2+2, \nu_3-2} \nonumber \\
&+(\nu_2-1)(\nu_3-1)\bra{\nu_1, \nu_2-2, \nu_3-2} \nonumber \\  
&+(\nu_1+1)(\nu_3+1)\bra{\nu_1+2, \nu_2, \nu_3+2} \nonumber \\ 
&-(\nu_1-1)(\nu_3+1)\bra{\nu_1-2, \nu_2, \nu_3+2} \nonumber \\ 
&-(\nu_1+1)(\nu_3-1)\bra{\nu_1+2, \nu_2, \nu_3-2} \nonumber \\
&+(\nu_1-1)(\nu_3-1)\bra{\nu_1-2, \nu_2, \nu_3-2} \big] . 
\end{align}
The Hilbert space preserved by the (\ref{diffC}) operator is given by
\begin{align} 
\mathcal{\tilde{H}}^{\rm Phy}_{\vec{\varepsilon}} &={\rm Span} \Big(\bra{\varepsilon_1 + 2n} \otimes \bra{\varepsilon_2 + 2m} \otimes \bra{\varepsilon_3 + 2k}: \nonumber \\  &n,m,k \in \mathbb{Z} \Big) 
\end{align}
and therefore we call it the physical Hilbert space. The parameters $\varepsilon_i \in [0,1]$. The physical scalar product in $\mathcal{\tilde{H}}^{\rm Phy}_{\vec{\varepsilon}}$ is defined by the kinematical one by restriction to the states $\psi \in \mathcal{\tilde{H}}^{\rm Phy}_{\vec{\varepsilon}}$.

\subsection{Connection representation}
In this section we investigate the quantum theory described in the previous section in (dual) $\eta^i$ connection representation. Let us begin with the following Fourier transform $\mathcal{\tilde{H}}^{\rm Phy}_{\vec{\varepsilon}} \to \mathcal{H}^{\rm Phy}_{\vec{\varepsilon}}=L^2(S^1, d\mu_{S^1})^{\otimes 3}$
\begin{align}\label{Fourier}
\bra{\tilde{\psi}} &= \sum_{n,m,k \in \mathbb{Z}} \psi(\varepsilon_1 + 2n, \varepsilon_2 + 2m, \varepsilon_3 + 2k) \times \nonumber \\ & \bra{\varepsilon_1 + 2n} \otimes \bra{\varepsilon_2 + 2m} \otimes \bra{\varepsilon_3 + 2k} \to \\  \psi(\eta^1, \eta^2, \eta^3) &=\sum_{n,m,k \in \mathbb{Z}} \psi(\varepsilon_1 + 2n, \varepsilon_2 + 2m, \varepsilon_3 + 2k) \times \nonumber \\ & e^{i\frac{\sqrt{\Delta}}{2}(\varepsilon_1 + 2n)\eta^1} e^{i\frac{\sqrt{\Delta}}{2}(\varepsilon_2 + 2m)\eta^2} e^{i\frac{\sqrt{\Delta}}{2}(\varepsilon_3 + 2k)\eta^3} \nonumber, 
\end{align}
where variables $\eta^i \in [0, 2\pi/\sqrt{\Delta}]$ \footnote{$L^2(S^1, d\mu_{S^1})^{\otimes 3}$ stands for the Hilbert space of square integrable functions on a $(S^1 \otimes S^1 \otimes S^1)$.} \footnote{The (\ref{Fourier}) is a 3 dimensional generalization of the Fourier transform introduced in \cite{KL-self} (see eq. (40) and (41) therein).}. From now on we will be interested in the most important case when $\varepsilon^i=0$ for $i=1,2,3$. The physical scalar product in the $\mathcal{H}^{\rm Phy}_{\vec{\varepsilon}=0}$ is defined in the obvious fashion:
\begin{equation}\label{scalarprod}
\ket{\psi} \bra{\phi} = \frac{(\sqrt{\Delta})^3}{(2\pi)^3} \int_0^{\frac{2\pi}{\sqrt{\Delta}}}\int_0^{\frac{2\pi}{\sqrt{\Delta}}}\int_0^{\frac{2\pi}{\sqrt{\Delta}}} d^3 \vec{\eta} \ \bar{\psi}(\vec{\eta}) \phi (\vec{\eta}) .
\end{equation}
It is not difficult to check that the basis element $$\ket{\vec{\eta}} \bra{2n,2m,2k} = e^{i\frac{\sqrt{\Delta}}{2} (2n)\eta^1} e^{i\frac{\sqrt{\Delta}}{2} (2m)\eta^2} e^{i\frac{\sqrt{\Delta}}{2}(2k)\eta^3}$$ is normalized to one with respect to (\ref{scalarprod}). It is easy to derive basic operators in the $\eta^i$ representation from the symplectic structure. Let us recall that the classical Poisson bracket is given by
\begin{equation}\label{poiss}
\{V_k, e^{i\frac{\bar{\mu}_j c^{(j)}}{2}} \}=-i 6 \pi G \gamma \sqrt{\Delta} l_{\rm Pl}^2 e^{i\frac{\bar{\mu}_j c^{(j)}}{2}} \delta^j_k \ ,
\end{equation}
where $\bar{\mu}_{(k)} c^k = \sqrt{\Delta} \eta^k$. After canonical quantization we get elementary operators. Components of the total volume operator are in the form
\begin{equation}\label{vol}
\hat{V}_k \psi(\eta^1,\eta^2,\eta^3)=-i 12 \pi \gamma l_{\rm Pl}^2 \frac{\partial}{\partial \eta^k} \psi(\eta^1,\eta^2,\eta^3)
\end{equation}
for $k=1,2,3$. 
The shift operators (\ref{shift}) act as multiplication operators
\begin{equation}
\widehat{e^{i\frac{\bar{\mu}_j c^{(j)}}{2}}}\psi(\eta^1,\eta^2,\eta^3)=e^{i\frac{\bar{\mu}_j c^{(j)}}{2}} \psi(\eta^1,\eta^2,\eta^3) \ .
\end{equation}
Let us now consider a part of (\ref{reg}), namely, $\frac{\sin{\bar{\mu}_{(i)} c^i}}{\bar{\mu}_i} p_i$. If we define the $\hat{O}_k$ operators as
\begin{equation}\label{Ok}
\hat{O}_k=  -i\frac{\sin(\sqrt{\Delta} \eta^k)}{\sqrt{\Delta}} \frac{\partial}{\partial \eta^k} -\frac{i}{2} \cos(\sqrt{\Delta} \eta^k) ,
\end{equation}
then the operators corresponding to $\sin(\bar{\mu}_{(i)} c^i) \frac{p_i}{\bar{\mu}_i}$ in the symmetric factor ordering are defined as $ 12\gamma \pi l_{\rm Pl}^2 \hat{O}_k $ for $k=1,2,3$  \footnote{Notice that $12\gamma \pi l_{\rm Pl}^2 \hat{O}_k = \underline{\hat{O}}_k$  in the volume representation.}. The gravitational part of the scalar constraint operator (\ref{scalar vol}) can now be written as
\begin{equation}\label{constr-eta}
\hat{C}_{\rm gr}=-18 \pi G \hbar^2  \left( \hat{O}_1 \hat{O}_2 +\hat{O}_2 \hat{O}_3+ \hat{O}_1 \hat{O}_3 \right).
\end{equation}
The operator (\ref{constr-eta}) is now symmetric with respect to the scalar product (\ref{scalarprod}), because the commutator $[\hat{O}_i, \hat{O}_j ]=0$ for $i \neq j$. Moreover, the spectrum of (\ref{constr-eta}) can be determined from the following equation
\begin{equation}
\hat{C}_{\rm gr} \ \psi_{(\vec{\lambda})}(\eta^1,\eta^2,\eta^3) = -18\pi G \hbar^2 \omega(\vec{\lambda}) \ \psi_{(\vec{\lambda})}(\eta^1,\eta^2,\eta^3).
\end{equation}
The eigenvalue has a form $\omega(\vec{\lambda})=\lambda_1 \lambda_2 + \lambda_1 \lambda_3 + \lambda_2 \lambda_3$ if the eigenfunction is given by $$\psi_{(\vec{\lambda})}(\vec{\eta})=\psi_{(\vec{\lambda})}(\eta^1,\eta^2,\eta^3)=\psi_{(\lambda_1)}(\eta^1) \psi_{(\lambda_2)}(\eta^2) \psi_{(\lambda_3)}(\eta^3) ,$$ where $\psi_{(\lambda_k)}(\eta^k)$ are defined by the formulas
\begin{equation}\label{eigenfun}
\psi_{(\lambda_k)}(\eta^k)=\frac{1}{2} \frac{\exp(i \lambda_k\ln |\tan \frac{\sqrt{{\Delta}} \eta^k}{2}|)}{\sqrt{|\sin \sqrt{\Delta} \eta^k|}}.
\end{equation}
for $k=1,2,3$. The spectrum of (\ref{constr-eta}) is continuous, because its eigenfunctions are normalized to Dirac delta $\ket{\psi_{(\vec{\lambda}^{\prime})}} \bra{\psi_{(\vec{\lambda})}} = \delta^3 (\vec{\lambda}^{\prime} - \vec{\lambda})$. Followed by the strategy adopted in \cite{5,APS} we can now write down the quantum constraint equation as 
\begin{equation}\label{equ-con}
-2\hbar^{-2} \hat{C}_{\rm gr} \psi(\eta^1, \eta^2, \eta^3, \phi)=-\partial^2_{\phi} \psi(\eta^1, \eta^2, \eta^3, \phi).
\end{equation}
and interpret the scalar field $\phi$ as time in the quantum theory. In order to do this, we can now decompose the Eq. (\ref{equ-con}) into  positive end negative frequencies as
\begin{equation}\label{evol}
\hat{\sqrt{|H|}} \psi(\eta^1, \eta^2, \eta^3, \phi)=\pm  \ i\partial_{\phi} \psi(\eta^1, \eta^2, \eta^3, \phi),
\end{equation}
where $\hat{H}=-2\hbar^{-2} \hat{C}_{\rm gr}$. The $\sqrt{|\hat{H}|}$ operator generates translations in the quantum "time" $\phi$ therefore we call it quantum Hamiltonian. The wave function which satisfies the two above equations can be written as
\begin{equation}\label{solut}
\Psi(\vec{\eta}, \phi_0) = \int_{\omega ( \vec{\lambda} ) > 0} d^3 \vec{\lambda} \ \tilde{\psi}(\vec{\lambda}) \ \psi_{(\vec{\lambda})}(\vec{\eta}) e^{\mp i\sqrt{36 \pi G} \sqrt{\omega} (\phi_0 - \phi^{\star} )}
\end{equation}
where $\tilde{\psi}(\vec{\lambda})$ is a profile of the wave packet, which will be determined later. At the end of this section please notice an important property of the above solution. The (\ref{solut}) is symmetric with respect to $\pi/\sqrt{\Delta}$ in each variable $\eta^i$ because it shares the same symmetry with (\ref{eigenfun}). Moreover, the states which satisfy (\ref{evol}) are called physical states, because of the Dirac quantization program which is applied here.

\section{The unitary transformation $W$}
\subsection{Isotropic $k=0$ model}
In the recent papers \cite{KL-self,4} it was shown that the differential relation 
\begin{equation}\label{y-iso}
\sqrt{\Delta} d\eta / dy = \sin (\sqrt{\Delta} \eta)
\end{equation}
leads to a very simple formula for the gravitational part of the scalar constraint operator, namely 
\begin{equation}
\hat{C}_{\rm gr}= - \frac{\partial^2}{\partial y^2},
\end{equation}
in \cite{4}, or
\begin{equation}
\hat{C}_{\rm gr}= - \frac{\partial^2}{\partial y^2} + {\rm potential \ \rm term}
\end{equation}
in \cite{KL-self}, where (up to $\sqrt{12\pi G}$ factor) $y=\ln |\tan \frac{\sqrt{{\Delta}} \eta}{2}|$ \footnote{Our variable $\sqrt{\Delta} \eta$ is denoted $\lambda b$ in \cite{4} and $x/2$ in \cite{KL-self}.}. The range of $\sqrt{\Delta} \eta$ is $[0,\pi/\sqrt{\Delta}]$, so we can forget about taking an absolute value $|\tan \frac{\sqrt{{\Delta}} \eta}{2}|=\tan \frac{\sqrt{{\Delta}} \eta}{2}$ because $\tan \frac{\sqrt{{\Delta}} \eta}{2} \ge 0$ . Notice that in (\ref{y-iso}) $\sin (\sqrt{\Delta} \eta) \ge 0$ as well. The interval $[0,\pi/\sqrt{\Delta}]$ is mapped to a real line $\mathbb{R}$. We will see however that in the Diagonal Bianchi I model the situation is different.

\subsection{Diagonal Bianchi I model}
A simple generalization of (\ref{y-iso}) for the quantum Bianchi I model yields the following differential relation:
\begin{equation}\label{diff-y}
\sqrt{\Delta} d\eta^k / dy^k = \sin (\sqrt{\Delta} \eta^k),
\end{equation}
which gives us the formulas
\begin{equation}\label{y}
y^k=\ln \left|\tan \frac{\sqrt{{\Delta}} \eta^k}{2}\right|.
\end{equation}
From (\ref{Fourier}) however we will find that the range of the variables $\eta^k$ is now $[0, 2\pi/\sqrt{\Delta}]$ for $k=1,2,3$ (in contrast to the isotropic case where the range is $[0,\pi/\sqrt{\Delta}]$). Let us consider this relation in more detail. Notice that: (1) $y^k(\eta^k)$ is not a bijection; (2) $y^k(\eta^k)$ is symmetric with respect to $\pi/\sqrt{\Delta}$. Moreover, from (\ref{y}) it is clear that the interval $[0, \pi/\sqrt{\Delta}]$ is mapped to a real line (denote it by $\mathbb{R}_{+}$) because $y^k(0)=-\infty$ and $y^k(\pi/\sqrt{\Delta})=\infty$. The second interval $[\pi/\sqrt{\Delta},2\pi/\sqrt{\Delta}]$ is mapped to another real line (denote it by $\mathbb{R}_{-}$) because $y^k(\pi/\sqrt{\Delta})=\infty$ and $y^k(2\pi/\sqrt{\Delta})=-\infty$. Now, because the right-hand side of (\ref{diff-y}) as well as $\tan \frac{\sqrt{{\Delta}} \eta^k}{2}$ in Eq. (\ref{y}) are positive for $\eta^k \in [0, \pi/\sqrt{\Delta}]$ and negative for $\eta^k \in [\pi/\sqrt{\Delta}, 2\pi/\sqrt{\Delta}]$ (for $k=1,2,3$), it is natural to consider $y^k(\eta^k)$ in each of the two intervals separately. 
Let us now consider the physical scalar product (\ref{scalarprod}) and the physical states (\ref{solut}) in more detail. Because of the very important symmetry of (\ref{solut}) with respect to $\pi/\sqrt{\Delta}$ in each of the variables $\eta^i$, the physical scalar product (\ref{scalarprod}) can be reduced to
\begin{equation}\label{sc-prod-symm}
\ket{\psi} \bra{\phi} = 8\frac{(\sqrt{\Delta})^3}{(2\pi)^3} \int_0^{\frac{\pi}{\sqrt{\Delta}}}\int_0^{\frac{\pi}{\sqrt{\Delta}}}\int_0^{\frac{\pi}{\sqrt{\Delta}}} d^3 \vec{\eta} \ \bar{\psi}(\vec{\eta}) \phi (\vec{\eta})
\end{equation}
Keeping this property in mind we will now construct a quantum theory in the $y^k$ variables in the symmetric sector for $\eta^k \in [0, \pi/\sqrt{\Delta}]$. Equations (\ref{diff-y}, \ref{y}) define a unitary map $W$ (which is a three-dimensional generalization of the map introduced in \cite{KL-self}) as
\begin{align}\label{mapa}
W(\psi)(y^1, y^2, y^3)&=(\sqrt{\Delta})^{3/2} \sqrt{\frac{d\eta^1}{dy^1}} \sqrt{\frac{d\eta^2}{dy^2}} \sqrt{\frac{d\eta^3}{dy^3}} \nonumber \\ &\times \psi(\eta^1(y^1),\eta^2(y^2),\eta^3(y^3)) 
\end{align}
and $W^{-1}$ as
\begin{align}\label{mapa1}
W^{-1}(f)(\eta^1,\eta^2,\eta^3)&= \frac{1}{(\sqrt{\Delta})^{3/2}}\sqrt{\frac{dy^1}{d\eta^1}}\sqrt{\frac{dy^2}{d\eta^2}}\sqrt{\frac{dy^3}{d\eta^3}} \nonumber \\ &\times f(y^1(\eta^1),y^2(\eta^2),y^3(\eta^3)).
\end{align}
Let us denote by $S^{1}_{(1/2)}$ a half of the circle, namely the interval $[0, \pi/\sqrt{\Delta}]$. The $W$ map transforms then the Hilbert space $L^2(S^{1}_{(1/2)} , d\mu_{S^1})^{\otimes 3}$ to the $L^2(\mathbb{R}^3,d^3\vec{y})$. The scalar product in the new Hilbert space is defined as
\begin{align}\label{sc-prod-y}
\ket{\psi} \bra{\phi}&=\frac{8}{(2\pi)^3} \int_{-\infty}^{\infty} \int_{-\infty}^{\infty} \int_{-\infty}^{\infty} dy^1 dy^2 dy^3 \nonumber \\ &\times W(\bar{\psi} )(y^1, y^2, y^3) W(\phi )(y^1, y^2, y^3) .
\end{align}
Let us now consider the operators. The $\hat{O}_k$ [defined by (\ref{Ok})] are transformed under $W$ to well-known operators as follows:
\begin{equation}
W (- i \frac{\sin(\sqrt{\Delta} \eta^k)}{\sqrt{\Delta}} \frac{\partial}{\partial \eta^k} - \frac{i}{2} \cos(\sqrt{\Delta} \eta^k)) W^{-1}= -i \frac{\partial}{\partial y^k} 
\end{equation}
for $\eta^k \in [0, \pi/\sqrt{\Delta}]$ (where $k=1,2,3$) only. The components of the volume operator (\ref{vol}) are transformed under $W$ as
\begin{align}\label{vol-y}
W \ \hat{V}_k \ W^{-1}&= 12 \pi \gamma l_{\rm Pl}^2 W \big( -i \frac{\partial}{\partial \eta^k} \big) W^{-1}= \\ &12 \pi \gamma l_{\rm Pl}^2 \sqrt{\Delta} \big(-i \cosh(y^k)\frac{\partial}{\partial y^k} -\frac{i}{2} \sinh(y^k) \big). \nonumber
\end{align}
Moreover it is not difficult to find a form of any physical state (\ref{solut}) in new variables
\begin{equation}\label{solut-y}
W(\Psi)(\vec{y}, \phi_0) = \int_{\omega ( \vec{\lambda} ) > 0} d^3 \vec{\lambda} \ \tilde{\psi}(\vec{\lambda}) \ e^{i\vec{\lambda} \cdot \vec{y} } e^{\mp i\frac{3}{\sqrt{2}}\sqrt{8 \pi G} \sqrt{\omega} (\phi_0 - \phi^{\star} )}
\end{equation}
where $\vec{\lambda} \cdot \vec{y} = \lambda_1 y^1 + \lambda_2 y^2 + \lambda_3 y^3$.

At the end of this subsection please notice the simple fact that if the physical states (\ref{solut}) share the following property $\Psi(\vec{\eta})=0$ at the boundary of the region of the integration in (\ref{sc-prod-symm}) then the operators (\ref{vol}) and (\ref{vol-y}) are symmetric on $L^2(S^{1}_{(1/2)} , d\mu_{S^1})^{\otimes 3}$ and $L^2(\mathbb{R}^3,d^3\vec{y})$ respectively.

\subsection{Classical and quantum evolution}
In this section we present the differences between classical and quantum evolution of the scale factors in the diagonal Bianchi type I model.  The classical relation between elementary variables and the scale factors [in the metric tensor (\ref{metric})] is 
\begin{align}
|\tilde{p}_1| &= a_2 a_3 \nonumber \\
|\tilde{p}_2| &= a_1 a_3 \\
|\tilde{p}_3| &= a_1 a_2 \ .\nonumber
\end{align}
New variables were introduced, namely $V_i = {\rm sqn} (p_i) |p_i|^{3/2}$, and these have corresponding operators in the quantum theory. The classical trajectories are defined as 
\begin{equation}\label{v-phi}
V_i(\phi) = V_{0i} \exp \Big(\pm \sqrt{8 \pi G} \frac{3}{2} \big( \frac{1 - \kappa_i}{|\kappa_{\phi}|} \big) (\phi - \phi_0 ) \Big) \ .
\end{equation}
Moreover, the parameters $\kappa_i$ and $\kappa_{\phi}$ satisfy relations (\ref{kappa}). In order to compare the classical and quantum model we can use the semiclassical states which are analogues of those introduced in \cite{5} \footnote{In \cite{5} the scalar constraint operator is defined by a different factor ordering in the volume representation, so the factor $(\phi_0 - \phi^{\star})$ in (\ref{semi-state}.) is multiplied by a different constant. See (8.41) in \cite{5}} \footnote{The state (\ref{semi-state}) is not normalized to one.}, namely
\begin{equation}\label{semi-state}
\Psi(\vec{\eta}, \phi_0) = \int_{\omega ( \vec{\lambda} ) > 0} d^3 \vec{\lambda} \ \tilde{\psi}(\vec{\lambda}) \ \psi_{(\vec{\lambda})}(\vec{\eta}) e^{\mp i\frac{3}{\sqrt{2}}\sqrt{8 \pi G} \sqrt{\omega} (\phi_0 - \phi^{\star} )}
\end{equation}
where the amplitude is given by
\begin{equation}\label{profile}
\tilde{\psi}(\vec{\lambda}) = e^{- \frac{(\lambda_1 - \mathcal{K} \kappa_1 )^2}{(2 \sigma_1 )^2}} e^{- \frac{(\lambda_2 - \mathcal{K} \kappa_2 )^2}{(2 \sigma_2 )^2}}  e^{- \frac{(\lambda_3 - \mathcal{K} \kappa_3 )^2}{(2 \sigma_3 )^2}} .
\end{equation}
Parameters $\kappa_i$ satisfy relations (\ref{kappa}). Consider now (\ref{semi-state}), which is sharply-peaked at some $\mathcal{K} \kappa_i$, for $i=1,2,3$. Since (\ref{profile}) is nonzero only in the small region close to $\vec{\lambda} = (\mathcal{K} \kappa_1, \mathcal{K} \kappa_2, \mathcal{K} \kappa_3)$ and moreover, if $\mathcal{K} \kappa_{i} \gg 1$ then $2(\lambda_1 \lambda_2 + \lambda_1 \lambda_3 + \lambda_2 \lambda_3) \approx \mathcal{K}^2 \kappa_{\phi}^2 > 0$ and we can approximate the integral (\ref{semi-state}) by
\begin{equation}\label{semi-state1}
\Psi(\vec{\eta}, \phi_0) = \int_{\mathbb{R}^3} d^3 \vec{\lambda} \ \tilde{\psi}(\vec{\lambda}) \ \psi_{(\vec{\lambda})}(\vec{\eta}) e^{\mp i\frac{3}{\sqrt{2}}\sqrt{8 \pi G} \sqrt{|\omega|} (\phi_0 - \phi^{\star} )} ,
\end{equation} 
because then the variables $\lambda_i$ ($i=1,2,3$) for which $\omega < 0$ do not contribute. For the symmetric sector defined by (\ref{sc-prod-symm}) the physical state (\ref{semi-state1}) in the $y_k$ variables has a well known form of Fourier transform in 3D
\begin{equation}\label{phy}
W(\Psi)(\vec{y},\phi_0)=\int_{\mathbb{R}^3} d^3 \vec{\lambda} \ \tilde{\psi}(\vec{\lambda}) e^{i \vec{\lambda} \cdot \vec{y}} e^{\mp i\frac{3}{\sqrt{2}}\sqrt{8 \pi G} \sqrt{|\omega|} (\phi_0 - \phi^{\star} )},
\end{equation}
with the amplitude (\ref{profile}). The integral (\ref{phy}) is not easy to compute analytically [apart from the case when $(\phi_0 - \phi^{\star}) = 0 $]. The difficulties come from the non linear behavior of the $|\omega(\vec{\lambda})|^{(1/2)}$. The function $e^{i \vec{\lambda} \cdot \vec{y}} \ \tilde{\psi}(\vec{\lambda}) e^{\mp i\frac{3}{\sqrt{2}}\sqrt{8 \pi G} \sqrt{\omega} (\phi_0 - \phi^{\star} )} $, however, is under good analytical control and we can compute the integral (\ref{phy}) numerically using fast Fourier transform in three-dimensions (see appendix for details). 

From the result of numerical simulations one can notice an important property of (\ref{phy}) and as a consequence of the $W$ map, the property of (\ref{semi-state}). Numerically computed $W(\Psi)(\vec{y},\phi_0)$ has a finite norm \footnote{A norm of (\ref{phy}) can be also computed analytically.} in the sense of the scalar product (\ref{sc-prod-y}). Moreover,  $W(\Psi)(\vec{y},\phi)$ decreases as $\exp(-a y_k^2)$ (where $a>0$ is some constant) in every direction $y^k$ for all values of $\phi$ used in the numerical simulations. If we use the inverse transformation $W^{-1}$ (\ref{mapa1}) and write the state (\ref{phy}) as
\begin{align}\label{zero-bound}
\Psi(\vec{\eta}, \phi) &= \frac{1}{\sqrt{\sin(\sqrt{\Delta} \eta^1)})} \frac{1}{\sqrt{\sin(\sqrt{\Delta} \eta^2)}} \frac{1}{\sqrt{\sin(\sqrt{\Delta} \eta^2)}} \nonumber \\ &\times W(\Psi)(\vec{y}(\vec{\eta}),\phi) ,
\end{align}
we conclude that $\Psi(\vec{\eta}, \phi) = 0$ at the boundary of the integration region in (\ref{sc-prod-symm}). In order to make it more clear, let us write the right-hand side of (\ref{zero-bound}) in terms of $y^k$ variables as
\begin{equation}
{\rm RHS} = \sqrt{\cosh y^1} \sqrt{\cosh y^2} \sqrt{\cosh y^3} W(\Psi)(\vec{y},\phi) .
\end{equation}
When $\eta^k \to 0$ from (\ref{y}) we get $y^k \to -\infty$, so the right-hand side and $\Psi(\vec{\eta}, \phi) \to 0$. Similarly, when $\eta^k \to \pi/\sqrt{\Delta}$ we get $y^k \to \infty$ and again $\Psi(\vec{\eta}, \phi) \to 0$, because 
\begin{equation}
\lim_{y^k \to \pm \infty} \sqrt{\cosh y^1} \sqrt{\cosh y^2} \sqrt{\cosh y^3} W(\Psi)(\vec{y},\phi) = 0 
\end{equation}
for $k=1,2,3$. This property is very nice, because then the operators (\ref{vol}) are symmetric on semiclassical states with respect to (\ref{sc-prod-symm}). Once the semiclassical states are known it is possible to compute the expectation values of the components of the volume operator (\ref{vol}) on semiclassical states, namely,
\begin{equation}\label{e-value}
\langle \hat{V}_k \rangle (\phi) = \frac{\ket{W(\Psi)}  \bra{W \ \hat{V}_k W^{-1} \ W(\Psi)}_{\phi}}{\ket{W(\Psi)} \bra{W(\Psi})}.
\end{equation}
The scalar product and the components of the volume operator are given by (\ref{sc-prod-y}) and (\ref{vol-y}) respectively. The above integrals were computed numerically (see appendix for details). 

\begin{widetext}

\begin{figure}[h]
\begin{center} 
\includegraphics[scale=0.55]{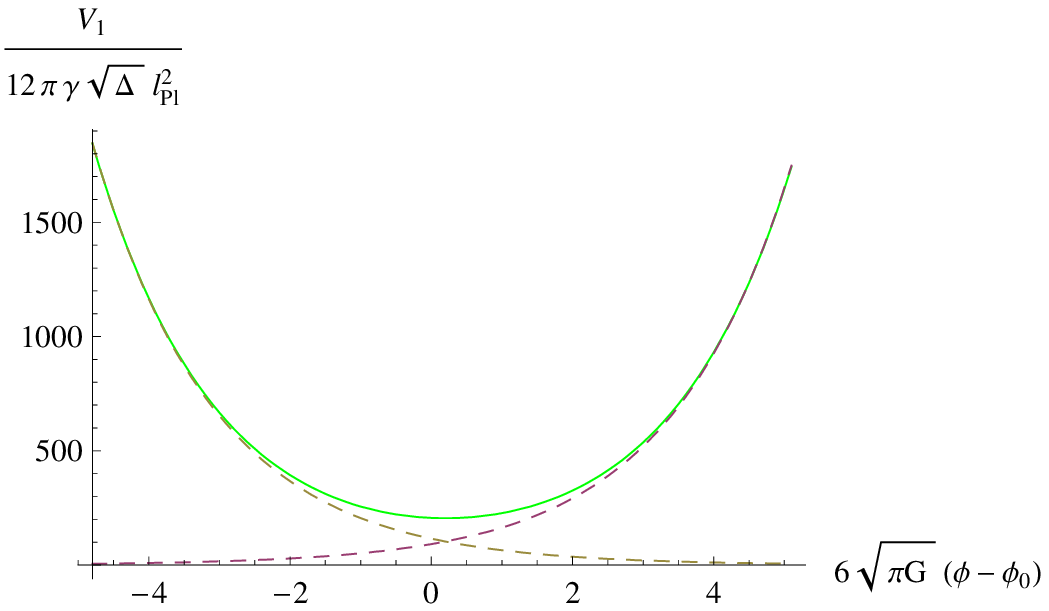}
\includegraphics[scale=0.55]{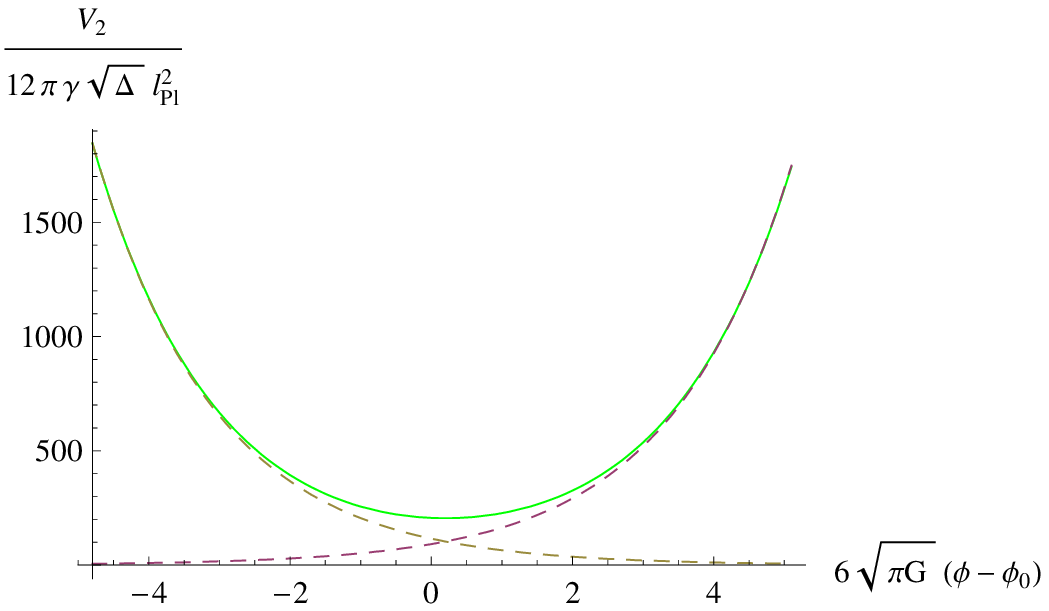}
\includegraphics[scale=0.55]{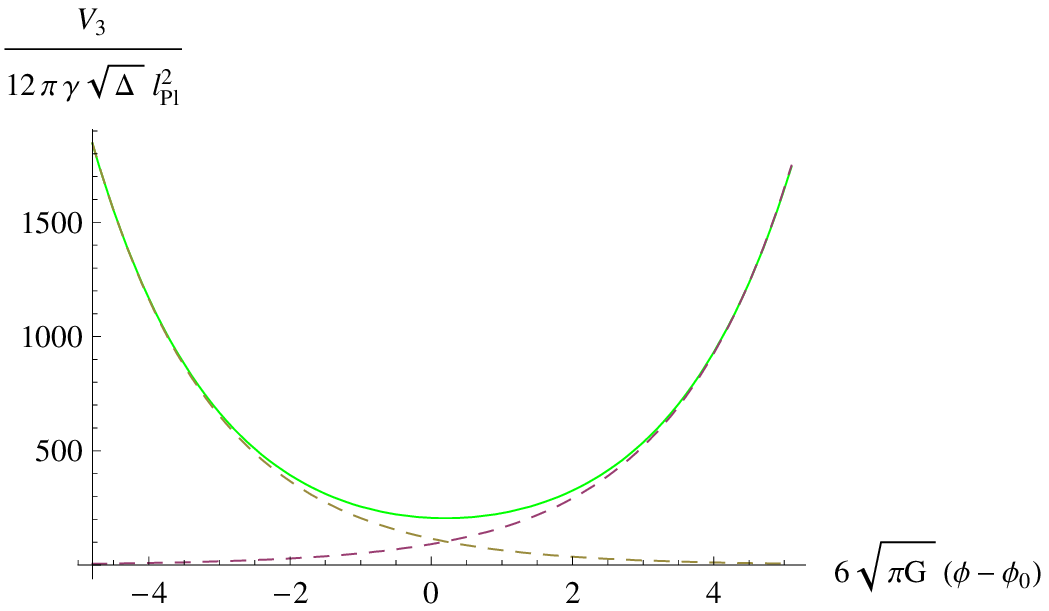}  
\end{center}
\caption{\textbf{Isotropic solution} for $\kappa_1=\kappa_2=\kappa_3=1/3$, $\kappa_{\phi}=\sqrt{2}/\sqrt{3}$. On the vertical axis $V_i$ denotes the expectation value $\langle \hat{V}_i \rangle(\phi)$ . The dashed lines denote classical trajectories.}\label{Fig1}
\begin{center}
\includegraphics[scale=0.55]{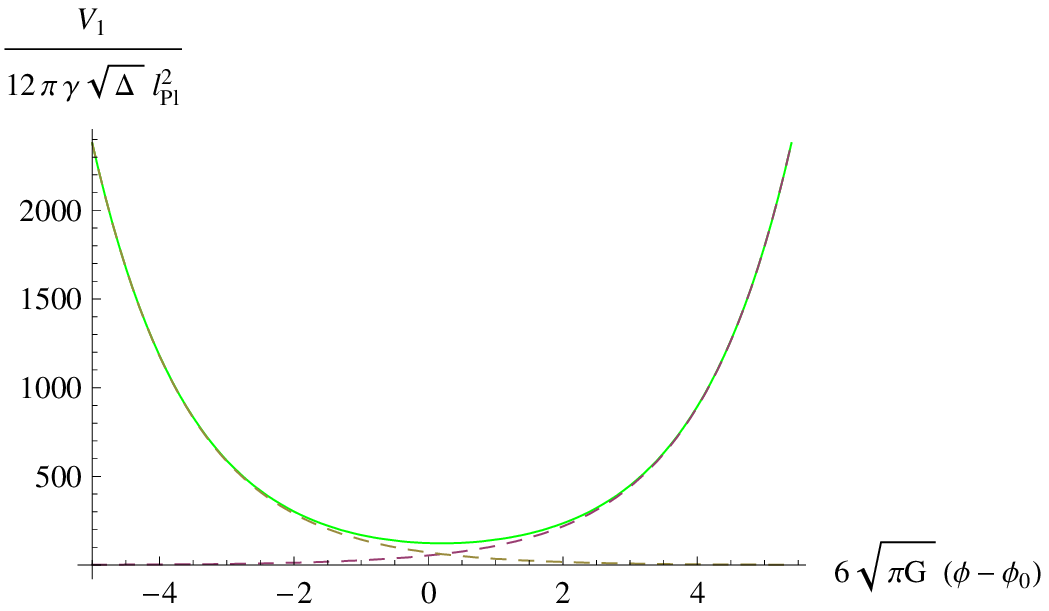}
\includegraphics[scale=0.55]{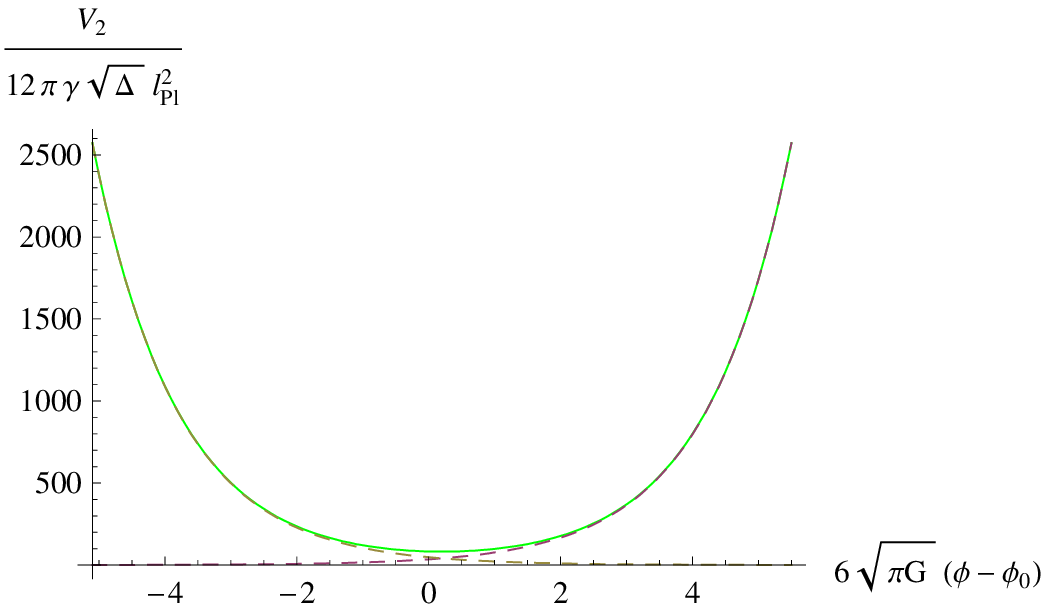}
\includegraphics[scale=0.55]{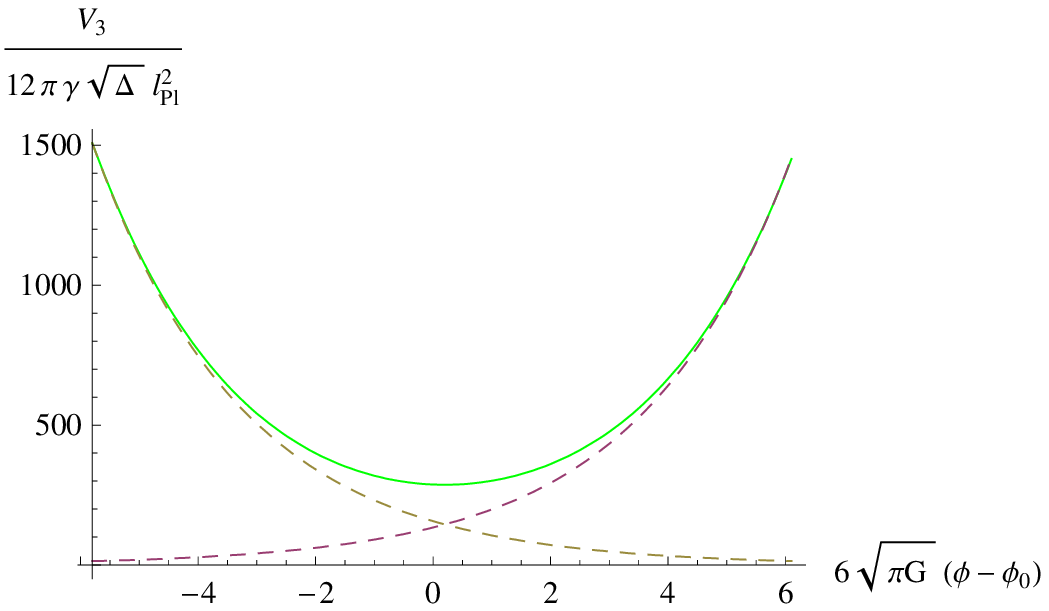}
\end{center}
\caption{\textbf{Kasner-unlike solution} for $\kappa_1 = 1/4$, $\kappa_2 = 1/6$, $\kappa_3 = 7/12,$ and $\kappa_{\phi} \approx 0.7546$. $V_i$ stands for the expectation values $\langle \hat{V}_i \rangle(\phi)$. The dashed lines denote classical trajectories.}\label{Fig2}
\begin{center}
\includegraphics[scale=0.55]{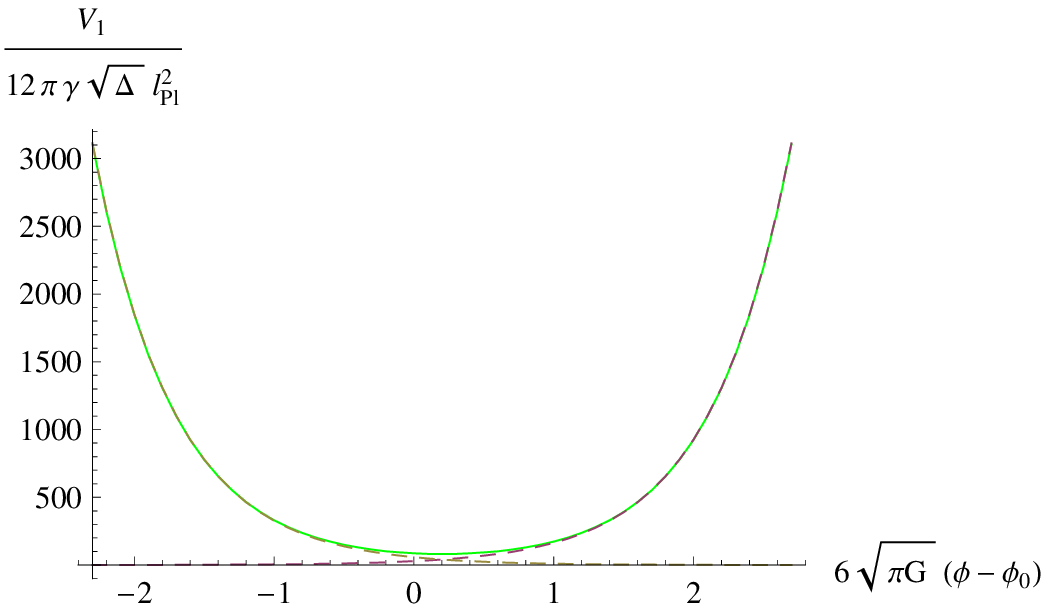}
\includegraphics[scale=0.55]{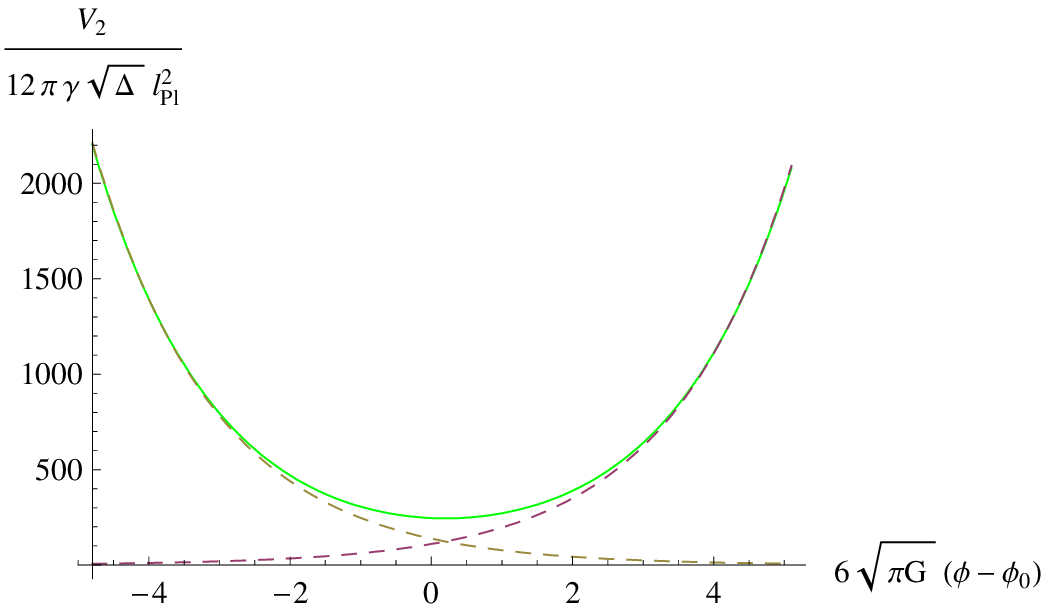}
\includegraphics[scale=0.55]{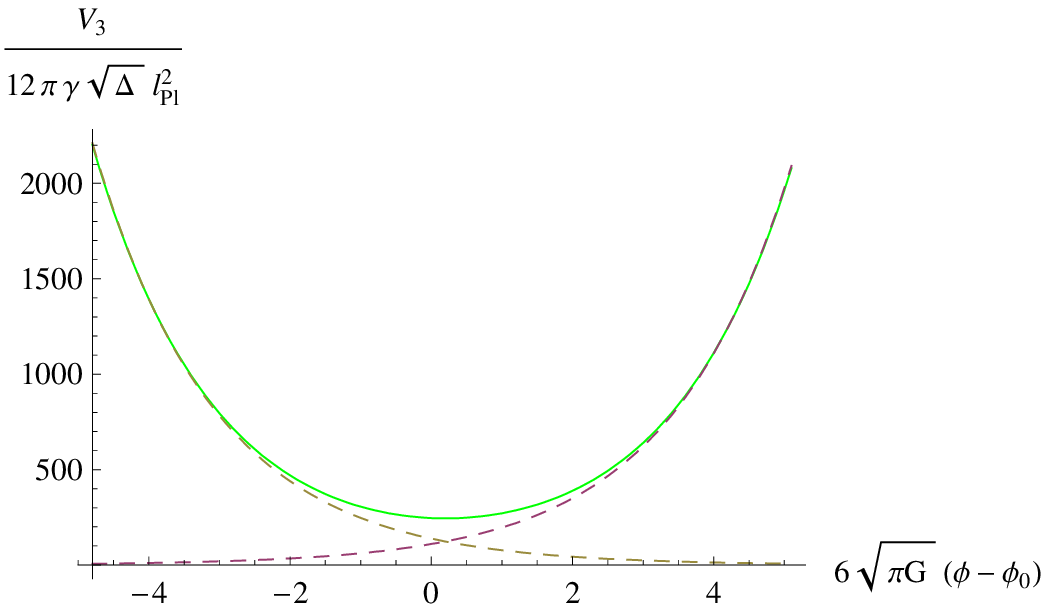}
\end{center}
\caption{\textbf{Kasner-like solution} for $\kappa_1 = -1/5$, $\kappa_2 = 3/5$, $\kappa_3 = 3/5,$ and $\kappa_{\phi} = \sqrt{6}/5$. Expectation values $\langle \hat{V}_i \rangle(\phi):=V_i$ are compared with the classical trajectories (dashed lines).}\label{Fig3}
\end{figure}

\end{widetext}

\section{Discussion}
\begin{figure}[h]
\begin{center}
\includegraphics[scale=0.6]{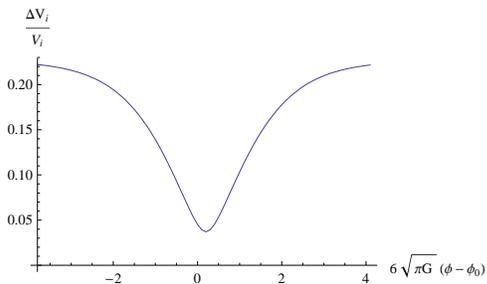}
\end{center}
\caption{The behavior of relative dispersion is plotted as a function of $\phi$ (This plot stands for isotropic solution, however for the Kasner-like case the plot is practically the same for all $V_i$.).}\label{Fig4}
\end{figure}

In the Figs. (\ref{Fig1}), (\ref{Fig2}) and (\ref{Fig3}) expectation values (\ref{e-value}) computed in different semiclassical states (\ref{phy}) are plotted as functions of the quantum time $\phi$. In order to make the pictures clear, each figure contains quantum and classical behavior of one degree of freedom $V_i$. Three solutions: isotropic, Kasner-like, and Kasner-unlike are compared with classical trajectories. In each case far away from classical singularity quantum trajectory follows closely the classical one up to the region where quantum gravitational effects become dominant. Instead of following the classical singular solution we have "big bounce" to another classical trajectory defined by (\ref{v-phi}). While it seems that classical singularity is avoided we will see later that unfortunately this result as well as the semi-classical limit of the theory does not have invariant meaning with respect to fiducial cell and fiducial metric. However, having now all types of numerical semiclassical solutions of (\ref{e-value}) we can make a phenomenological observation. Each curve in Figs. (\ref{Fig1},\ref{Fig2},\ref{Fig3}) can be well approximated by the following hyperbolic function:
\begin{equation}
\langle \hat{V}_i \rangle(\phi) = V_{0i} \cosh \big( \sqrt{8 \pi G} \frac{3}{2} \big( \frac{1 - \kappa_i}{|\kappa_{\phi}|} \big) (\phi - \phi^0 ) \big) \ ,
\end{equation}
where values $V_{0i}$ and $\phi^0$ depend on semiclassical states $\psi$ which are used in computations. Now, we consider numerical parameters used in calculations and dispersions defined by $$\Delta V_i = \sqrt{\langle \hat{V}_i^2 \rangle - \langle \hat{V}_i \rangle^2} \ .$$ Typical values of $\mathcal{K}\kappa_i$ in the Fourier amplitude (\ref{profile}) used in numerical calculations are as follows: $\mathcal{K}\kappa_i \approx 100, 150, 200, 250, 350, 450$. Typical behavior of dispersions is plotted in Fig. (\ref{Fig4}) for isotropic and Kasner-unlike solutions. As can be seen, before and after big bounce relative dispersions $\Delta V_i / \langle \hat{V}_i \rangle$ are asymptotically constant. Typical asymptotic values are $\Delta V_i / \langle \hat{V}_i \rangle \approx 0.2$ in our numerical calculations.
The integral (\ref{e-value}) and dispersions were computed using lattices defined by parameters $N_1=N_2=N_3=256, 320, 512$ [see (\ref{3int})]. In principle one can use better parameters $N_i$ in (\ref{3int}) and higher values of $\mathcal{K} \kappa_i$ to increase the accuracy and semiclassicality of the calculations. However, the requirements for the calculating machine are beyond the parameters of the "SOWA" cluster in the University of Warsaw physics department which the author used and whose capability allows for maximum values of $N_1$, $N_2$, and $N_3$  in (\ref{3int},\ref{sc-num}) equations to be 512. Let us now mention the difficulties with the Kasner-like solution. Computations of small dispersions for this case are unfortunately impossible to perform for the author. This is due to the fact that in order to have a good approximation in (\ref{semi-state}) $$\int_{\omega ( \vec{\lambda} ) > 0} \to \int_{\mathbb{R}^3}$$ defined by $2(\lambda_1 \lambda_2 + \lambda_1 \lambda_3 + \lambda_2 \lambda_3) \approx \mathcal{K}^2 \kappa_{\phi}^2 > 0$ for sufficiently large $\sigma_i$ parameters, we have to take large numbers $\mathcal{K} \kappa_i$, which is not possible according to the random access memory (RAM) capability of the "SOWA" cluster. Our maximal lattice defined by $N_i=512$ ($i=1,2,3$) is not dense enough to compute integral (\ref{e-value}) with very large values of $\mathcal{K}\kappa_i$. However, we do not see a reason why dispersions for the Kasner-like solution would be different than in the Kasner-unlike solution, so we expect very similar behavior.

\begin{figure}[h]
\begin{center}
\includegraphics[scale=0.6]{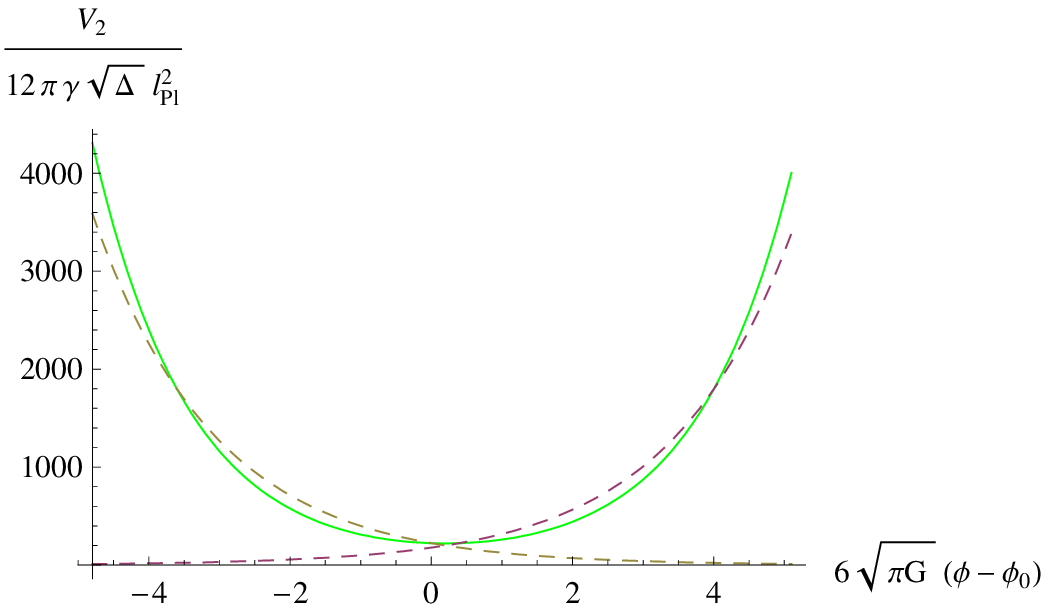}
\includegraphics[scale=0.6]{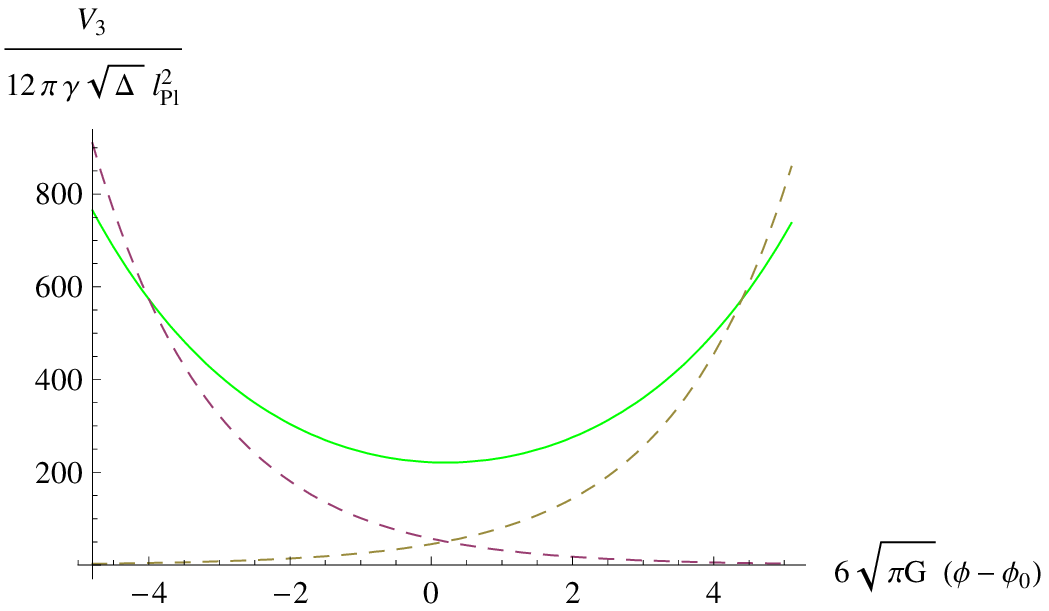}
\end{center}
\caption{The example of the isotropic solutions $\langle \hat{V}_k \rangle$ (for $k=2,3$) plotted as a function of $\phi$. The fiducial cell is rescaled in one direction as $V_0 = l_1 l_2 l_3 = l^3 \to V_0=\lambda_1 l_1 l_2 l_3 = \lambda_1 l^3$, where $\lambda_1=2$.}\label{Fig5}
\end{figure}

Results enclosed in this paper show only semiclassical behavior of the three gravitational degrees of freedom. However, the above results are valid only for the cubical shape of the fiducial cell and metric. Moreover only the cubical scaling of the theory is allowed. Let us now focus on the scaling properties of the gravitational part of the scalar constraint. If we allow one to scale $a_i \to l_i a_i$, where all $l_i$ are different for $i=1,2,3$ (the fiducial cell becomes cuboid) then the spectrum of the operator (\ref{scalar vol}) scales as
\begin{equation}
\omega \sim \frac{1}{V_0^2} (\alpha_1 \lambda_1 \lambda_2 + \alpha_2 \lambda_1 \lambda_3 +\alpha_3 \lambda_2 \lambda_3) 
\end{equation}
where $V_0=l_1 l_2 l_3$ and $\alpha_1, \alpha_2, \alpha_3$ are some numbers determined by $l_1, l_2, l_3$. For $l_i \neq 1$ we have $\alpha_i \neq 1$ and then semiclassical states (\ref{semi-state}) do not give correct classical limit. The quantum trajectory defined by (\ref{e-value}) simply does not follow the classical one as one can see in the Fig. \ref{Fig5}.

Thus the semiclassical limit with respect to (\ref{semi-state}) is fiducial cell dependent. Moreover, the shift operators defined by $\sin\bar{\mu}_i c^i$ correspond to the different steps after rescaling $V_0 \to V_0l_1 l_2 l_3$, in contrast to the $k=0$ model \cite{APS}, where the step defined by the $\sin\bar{\mu}c$ is invariant under $V_0 \to l^3 V_0$. One can conclude that for each choice of the (noncubical) fiducial cell $\mathcal{V}$ (and fiducial metric) the model has different quantum dynamics, while it seems reasonable to expect that the quantum model should generate only one physical dynamics. The only consistent fiducial cell is a cube with the rescaling of the scale factors $a_i \to l a_i$. Then $\alpha_1=\alpha_2=\alpha_3=V_0^2$, the spectrum is invariant $\omega \to \omega$, and the semiclassical limit defined by (\ref{semi-state}) is correct. 

Let us conclude that the $\bar{\mu}$ quantization scheme which is highly successful in isotropic models, applied for the Bianchi I in \cite{5} has its problems described above. The classical theory allows one to consider many different fiducial cells, while the quantum model allows only the cubical one. Moreover, the fiducial metric $$^{\textrm{\tiny{0}}} q_{ab} = \ \bazad{1}{a} \ \bazad{1}{b} + \bazad{2}{a} \ \bazad{2}{b} + \bazad{3}{a} \ \bazad{3}{b}$$ also must be isotropic, which is a serious limitation. Although for the 3-Torus topology of the spatial slice $\Sigma$ with fixed coordinate volume $V_0$ the $\bar{\mu}$ scheme can make sense \cite{MGM,MMP}, for the noncompact $\Sigma$ topology the model is tentative and should be considered as such. 

Now what about future directions for the given model? It seems necessary to carry out the analysis involving the curvature invariants in order to investigate the scaling properties of the model, especially at the bounce. However, our results suggest that the $\bar{\mu}$ scheme applied to the BI model does not have correct scaling properties even at the bounce. We conclude that the new way of loop regularization and its quantization is needed. Also similar problems may arise in a large class of models based on arbitrary discretizations such as lattice refinement models, so one should be more careful with respect to the regularization of the constraints. 

\section{Acknowledgments}
The author would like to thank I$\tilde{\rm n}$aki Garay, Wojciech Kami\'nski, Jerzy Lewandowski, Tomasz Pa{\l}owski and Tomasz Werner for useful discussions and especially W{\l}odzimierz Natorf for his help with computers. This work was partially supported by the Polish Ministry of Science and Higher Education, Grant No. N N202 007734 and Foundation for Polish Science, Master grant. 

\section{Appendix}
\subsection{Numerical Integration}
Consider a Fourier integral 
\begin{equation}\label{integ}
\psi(x) = \int_{-\infty}^{\infty} d \lambda \ \tilde{\psi}(\lambda) e^{i \lambda x} ,
\end{equation}
where the amplitude is sharply peaked at some value of $\lambda_0$ and zero everywhere else. In order to deal with the above integral numerically we can use the following strategy. Let us approximate (\ref{integ}) by
\begin{equation}\label{approx}
\psi(x) \approx \int_{a}^{b} d \lambda \ \tilde{\psi}(\lambda) e^{i \lambda x}.
\end{equation}
If the interval $[a,b]$ covers the region, where the $\tilde{\psi}(\lambda)$ is nonzero, and if this interval is large enough, then the (\ref{approx}) is a good approximation of the integral (\ref{integ}). The next step is to approximate the integral by defining a lattice on the $x$ axis $x_k = x_0 + k \Delta_x$ and on $\lambda$ axis $\lambda_j=a + j \Delta$, where $k,j=0,1,2...N-1$. Moreover, $\Delta_x = 2\pi /(b-a)$ and $\Delta = (b-a)/N$. The integral can be then approximated by the sum as 
\begin{align}
\psi(x_k) &\approx \Delta \sum_{j=0}^{N-1} \tilde{\psi}(\lambda_j ) e^{i \lambda_j x_k} \\ &= \Delta e^{i a x_0} e^{i a k \Delta_x} \sum_{j=0}^{N-1} e^{i j \Delta x_0} \tilde{\psi}(\lambda_j ) e^{i \frac{2\pi j k}{N}} \ \nonumber . 
\end{align}
The sum $X_k:=\sum_{j=0}^{N-1} \bar{\psi}_j e^{i \frac{2\pi j k}{N}}$, where $\bar{\psi}_j=e^{i j \Delta x_0} \tilde{\psi}(\lambda_j )$ can be computed using the powerful fast Fourier transform. As a result we get values of $\psi(x)$ on a lattice defined by $x_k = x_0 + k \Delta_x$, for $k=0,1,2,3...N-1$. For sufficiently large $N$ (and small $\Delta$ , $\Delta_x$) it is possible to cover the region on the $x$ axis where the function $\psi(x)$ is nonzero with a good accuracy. A norm of (\ref{integ}) can be computed as
\begin{align}
I &= \int_{-\infty}^{\infty} d x \ \bar{\psi}(x) \psi (x) \approx \int_{x_0}^{x_{N-1}} d x \ \bar{\psi}(x) \psi (x) \nonumber \\  &\approx \Delta_x \sum_{k=0}^{N-1} \bar{\psi}(x_k) \psi(x_k) \ .
\end{align}
It is now not difficult to develop a three-dimensional analogue. Let us consider the following 3D Fourier transform 
\begin{align}
\psi(y^1, y^2, y^3) &= \int_{-\infty}^{\infty} \int_{-\infty}^{\infty} \int_{-\infty}^{\infty} d \lambda_1 \ d \lambda_2 \ d \lambda_3  \nonumber \\ &\times \tilde{\psi}(\lambda_1, \lambda_2,\lambda_3) e^{i (\lambda_1 y^1 + \lambda_2 y^2 + \lambda_3 y^3)} \ .
\end{align}
If again the Fourier profile $\tilde{\psi}(\lambda_1, \lambda_2,\lambda_3)$ is sharply distributed, the integral can be approximated by 
\begin{align}
\psi(y^1, y^2, y^3) & \approx \int_{a_1}^{b_1} \int_{a_2}^{b_2} \int_{a_3}^{b_3} d \lambda_1 \ d \lambda_2 \ d \lambda_3  \nonumber \\ &\times \tilde{\psi}(\lambda_1, \lambda_2,\lambda_3) e^{i (\lambda_1 y^1 + \lambda_2 y^2 + \lambda_3 y^3)} \ ,
\end{align}
if the boundary of the integration region defined by $a_i$ and $b_i$ (for $i=1,2,3$) is appropriately chosen. The next step is to approximate the above integral by the sum [at some point $(y^1_{k_1},y^2_{k_2},y^3_{k_3})$] as
\begin{align}\label{3int}
\psi(y^1_{k_1},y^2_{k_2},y^3_{k_3}) & \approx \Delta_{1} \Delta_{2} \Delta_{3} \sum_{j_1=0}^{N_1-1} \sum_{j_2=0}^{N_2-1} \sum_{j_3=0}^{N_3-1}  \\ 
& \times \tilde{\psi}(\lambda_1^{j_1}, \lambda_2^{j_2},\lambda_3^{j_3}) e^{i (\lambda_1^{j_1} y^1_{k_1} + \lambda_2^{j_2} y^2_{k_2} + \lambda_3^{j_3} y^3_{k_3})} \nonumber .
\end{align}
If we set a lattice: 1) $y^1_{k_1} = y_0^1 + k_1 \Delta_{y_1}$, $y^2_{k_2} = y_0^2 + k_2 \Delta_{y_2}$ and $y^3_{k_3} = y_0^3 + k_3 \Delta_{y_3}$; 2) $\lambda_1^{j_1} = a_1 + \Delta_1 j_1$, $\lambda_2^{j_2} = a_2 + \Delta_2 j_2$ and $\lambda_3^{j_3} = a_3 + \Delta_3 j_3$, where $\Delta_1=(b_1 - a_1)/N_1 $, $\Delta_2=(b_2 - a_2)/N_{2}$, $\Delta_3=(b_3 - a_3)/N_3$, $\Delta_{y_1} = 2\pi/(b_1 - a_1)$, $\Delta_{y_2} = 2\pi/(b_2 - a_2)$ and $\Delta_{y_3} = 2\pi/(b_3 - a_3)$ then (\ref{3int}) has the following form:
\begin{equation}
\psi(y^1_{k_1},y^2_{k_2},y^3_{k_3}) \approx \Delta_{1} \Delta_{2} \Delta_{3} \ e^{i{\rm arg}(k_1,k_2,k_3)} X_{k_1, k_2, k_3} ,
\end{equation}
where ${\rm arg}(k_1,k_2,k_3) = y_0^1 a_1 + y_0^2 a_2 + y_0^3 a_3 + \Delta_{y_1} k_1 a_1 + \Delta_{y_2} k_2 a_2 + \Delta_{y_3} k_3 a_3$ and 
\begin{align}\label{big-sum}
X_{k_1, k_2, k_3} &= \sum_{j_1=0}^{N_1-1} \sum_{j_2=0}^{N_2-1} \sum_{j_3=0}^{N_3-1} \tilde{\psi}(\lambda_1^{j_1}, \lambda_2^{j_2},\lambda_3^{j_3}) \\ &\times e^{i(y_0^1 \ j_1 \ \Delta_1 \ + \ y_0^2 \ j_2 \ \Delta_2 \ + \ y_0^3 \ j_3 \ \Delta_3)} \nonumber \\ & \times \exp\Big[i 2\pi \big( \frac{k_1 j_1}{N_1} + \frac{k_2 j_2}{N_2} + \frac{k_3 j_3}{N_3}\big) \Big]  \nonumber ,
\end{align}
where $k_i = 0,1,2...N_i-1$, for $i=1,2,3$. Similarly to the one-dimensional case the sum (\ref{big-sum}) can be computed using fast Fourier transform in 3D. The scalar product (\ref{sc-prod-y}) is defined as 
\begin{align}\label{sc-num}
\ket{\psi} \bra{\psi} &\approx \frac{8}{(2\pi)^3}\Delta_{y_1} \Delta_{y_2} \Delta_{y_3} \sum_{k_1=0}^{N_1-1} \sum_{k_2=0}^{N_2-1} \sum_{k_3=0}^{N_3-1}  \nonumber \\ & \times 
\bar{\psi}(y^1_{k_1},y^2_{k_2},y^3_{k_3}) \psi(y^1_{k_1},y^2_{k_2},y^3_{k_3}) \ .
\end{align}

\end{document}